\documentclass[11pt,preprintnumbers,titlepage,nofootinbib]{revtex4-2}%
\usepackage[latin9]{inputenc}
\usepackage{amsmath}
\usepackage{amssymb}
\usepackage{graphicx}
\usepackage{esint}
\usepackage[unicode=true,  bookmarks=false,  breaklinks=false,pdfborder={0 0
1},backref=section,colorlinks=false]{hyperref}
\usepackage{textcomp}
\usepackage{color}
\usepackage{textcomp}
\usepackage{wasysym}
\usepackage{array}
\usepackage{multirow}
\usepackage{wasysym}
\usepackage{wasysym}
\usepackage{amsfonts}
\usepackage{subfigure}
\usepackage{cleveref}
\usepackage{listings}
\usepackage{xcolor}
\usepackage{array}
\usepackage{url}
\usepackage{color}
\usepackage{subfigure}
\usepackage{appendix}%
\hypersetup{
colorlinks,linkcolor=blue,citecolor=blue,urlcolor=blue}
\makeatletter
\DeclareFontEncoding{LGR}{}{}

\ProvideTextCommand{\~}{LGR}[1]{\char126#1}
\DeclareTextSymbolDefault{\textquotedbl}{T1}

\@ifundefined{textcolor}{}{
\definecolor{BLACK}{gray}{0}
\definecolor{WHITE}{gray}{1}
\definecolor{RED}{rgb}{1,0,0}
\definecolor{GREEN}{rgb}{0,1,0}
\definecolor{BLUE}{rgb}{0,0,1}
\definecolor{CYAN}{cmyk}{1,0,0,0}
\definecolor{MAGENTA}{cmyk}{0,1,0,0}
\definecolor{YELLOW}{cmyk}{0,0,1,0}
\definecolor{ballblue}{rgb}{0.13, 0.67, 0.8}
\definecolor{bleudefrance}{rgb}{0.19, 0.55, 0.91}
\definecolor{blue(ncs)}{rgb}{0.0, 0.53, 0.74}
\definecolor{darkpastelgreen}{rgb}{0.01, 0.75, 0.24}
\definecolor{darkspringgreen}{rgb}{0.09, 0.45, 0.27}
\definecolor{denim}{rgb}{0.08, 0.38, 0.74}
\definecolor{electricviolet}{rgb}{0.56, 0.0, 1.0}
}

\makeatother
\begin{document}
\preprint{CTP-SCU/2025013}
\title{Spin-induced Scalarized Black Holes in Einstein-Maxwell-scalar Models}
\author{Lang Cheng$^{a}$}
\email{chenglang@stu.scu.edu.cn}
\author{Guangzhou Guo$^{b}$}
\email{guogz@sustech.edu.cn}
\author{Peng Wang$^{a}$}
\email{pengw@scu.edu.cn}
\author{Haitang Yang$^{a}$}
\email{hyanga@scu.edu.cn}
\affiliation{$^{a}$College of Physics, Sichuan University, Chengdu, 610064, China}
\affiliation{$^{b}$Department of Physics, Southern University of Science and Technology,
Shenzhen, 518055, China}

\begin{abstract}
We construct spin-induced scalarized black hole solutions in a class of
Einstein-Maxwell-scalar models, where a scalar field is non-minimally coupled
to the electromagnetic field. Our results show that scalar hair develops only
for rapidly rotating black holes, while slowly spinning ones remain well
described by the Kerr-Newman (KN) metric. The scalar field contributes only a
small fraction of the total mass, indicating suppressed nonlinear effects.
This suppression may account for the narrow existence domains of scalarized
black holes and the similarities observed in their existence domains across
different coupling functions. Moreover, scalarized black holes are found to
coexist with linearly stable, entropically favored KN black holes. These
results motivate further investigations into the nonlinear dynamics and
stability of scalarized black holes in these models.

\end{abstract}
\maketitle
\tableofcontents

{}

{}

{}

\section{Introduction}

Recently, the LIGO and Virgo Collaborations reported the detection of
gravitational waves from binary black hole mergers
\cite{LIGOScientific:2016aoc}, while the Event Horizon Telescope Collaboration
produced the first images of the supermassive black holes M87{*} and Sgr A{*}
\cite{Akiyama:2019cqa,Akiyama:2019brx,Akiyama:2019sww,Akiyama:2019bqs,Akiyama:2019fyp,Akiyama:2019eap,Akiyama:2021qum,Akiyama:2021tfw,EventHorizonTelescope:2022xnr,EventHorizonTelescope:2022vjs,EventHorizonTelescope:2022wok,EventHorizonTelescope:2022exc,EventHorizonTelescope:2022urf,EventHorizonTelescope:2022xqj}%
. These groundbreaking observations have opened new avenues for testing
general relativity in the strong-field regime, including the no-hair theorem,
which asserts that stationary black holes are uniquely characterized by their
mass, angular momentum and charge
\cite{Israel:1967wq,Carter:1971zc,Ruffini:1971bza}. However, counterexamples
to the no-hair theorem---commonly referred to as \textquotedblleft
hairy\textquotedblright\ black holes---have been found in various theories,
where black holes possess additional degrees of freedom. The first such
solutions were discovered in the Einstein-Yang-Mills theory
\cite{Volkov:1989fi,Bizon:1990sr,Greene:1992fw}. Subsequently, black hole
solutions with Skyrme hair \cite{Luckock:1986tr,Droz:1991cx} and dilaton hair
\cite{Kanti:1995vq} were also obtained. For a comprehensive review, see
\cite{Herdeiro:2015waa}.

Recently, the no-hair theorem has been challenged by extended theories of
gravity that introduce non-minimal couplings between scalar fields and other
fundamental fields. A prominent mechanism for circumventing this theorem is
spontaneous scalarization, wherein a trivial scalar field configuration around
a black hole becomes unstable and evolves into a non-trivial configuration,
leading to the emergence of scalar hair
\cite{Damour:1993hw,Damour:1996ke,Cardoso:2013opa,Cardoso:2013fwa}. This
phenomenon is typically driven by a tachyonic instability, in which scalar
field perturbations acquire an effective negative mass squared on the black
hole background, triggering growth of the scalar field. In particular, it has
been demonstrated that in Einstein-scalar-Gauss-Bonnet (EsGB) theories, a
scalar field appropriately coupled to the Gauss-Bonnet invariant can develop a
tachyonic instability near Kerr black holes when the spin exceeds a critical
threshold \cite{Dima:2020yac}. This spin-induced instability has subsequently
been shown to lead to the formation of spin-induced scalarized black holes at
sufficiently high spins \cite{Herdeiro:2020wei,Berti:2020kgk}.

To gain deeper insight into the dynamical evolution of scalarization, a
technically simpler class of models known as Einstein-Maxwell-scalar (EMS)
models has been proposed~\cite{Herdeiro:2018wub}. In these models, a scalar
field is non-minimally coupled to the electromagnetic sector, enabling
scalarized black hole solutions that go beyond Kerr-Newman (KN) solutions
\cite{Fernandes:2019rez,Fernandes:2019kmh,Blazquez-Salcedo:2020nhs,Wang:2020ohb,Guo:2021zed,Guo:2023mda,Belkhadria:2023ooc}%
. Remarkably, scalarized black holes in the EMS models can exhibit multiple
light rings in the equatorial plane \cite{Gan:2021xdl,Guo:2023mda}. This
feature not only leads to distinctive optical signatures in black hole imaging
\cite{Gan:2021pwu,Guo:2022muy,Chen:2022qrw,Chen:2023qic,Chen:2024ilc} but is
also associated with the existence of long-lived quasinormal modes
\cite{Cardoso:2014sna,Keir:2014oka,Guo:2021bcw,Guo:2021enm,Guo:2022umh} and
potential superradiant instabilities \cite{Guo:2023ivz}.

While most studies of the EMS models have focused on positive
scalar-electromagnetic coupling constants---where scalarization is typically
driven by electric charge---recent work has shown that, for sufficiently
negative coupling constants, rapidly rotating KN black holes can also undergo
spontaneous scalarization triggered by a spin-induced instability
\cite{Hod:2022txa,Lai:2022ppn}. Furthermore, the domain of existence for
scalar clouds arising from this spin-induced instability has been thoroughly
investigated in the KN black hole parameter space \cite{Guo:2024lck}. These
scalar clouds are generally interpreted as signaling the onset of spin-induced
scalarized KN black holes.

In this paper, we numerically construct spin-induced scalarized KN black hole
solutions in the EMS models. The structure of the paper is as follows. In Sec.
\ref{sec:Set-Up}, we review the EMS models and describe numerical methods used
to obtain scalarized black hole solutions. Sec. \ref{sec:Numerical-Results}
presents numerical results for the spin-induced scalarized KN black holes. A
summary of our findings is given in Sec. \ref{sec:Conclusions}. Finally,
Appendix \ref{sec:Appendix-A:-Convergence Results for Spectral Method}
provides a convergence analysis of the numerical solutions, and Appendix
\ref{sec:Appendix-B:-Domain of Existence for Quadratic Coupling Function}
explores the domain of existence for spin-induced scalarized KN black holes
with a quadratic coupling function. Throughout this work, we adopt the
convention $G=c=4\pi\epsilon_{0}=1$.

\section{Setup}

\label{sec:Set-Up}

In this section, we first review the EMS models that exhibit a tachyonic
instability in KN black holes. We then present the numerical method used to
construct scalarized KN black hole solutions within these models. \ 

\subsection{Einstein-Maxwell-scalar Models}

In the EMS models, a scalar field is non-minimally coupled to
electromagnetism, potentially inducing a tachyonic instability in KN black
holes. The action for this system is
\begin{equation}
S=\frac{1}{16\pi}\int d^{4}x\sqrt{-g}\left[  R-2\partial_{\mu}\phi
\partial^{\mu}\phi-f\left(  \phi\right)  F^{\mu\nu}F_{\mu\nu}\right]  ,
\label{eq:Action}%
\end{equation}
where $R$ is the Ricci scalar, $\phi$ is the scalar field, and $F_{\mu\nu
}=\partial_{\mu}A_{\nu}-\partial_{\nu}A_{\mu}$ denotes the electromagnetic
field tensor. Varying the action $\left(  \ref{eq:Action}\right)  $ yields the
equations of motion for the metric $g_{\mu\nu}$, scalar field $\phi$ and
electromagnetic field $A_{\mu}$:
\begin{align}
R_{\mu\nu}-\frac{1}{2}Rg_{\mu\nu}  &  =2T_{\mu\nu},\nonumber\\
\square\phi-\frac{1}{4}\frac{df\left(  \phi\right)  }{d\phi}F^{\mu\nu}%
F_{\mu\nu}  &  =0,\label{eq:nonlinearEOMs}\\
\partial_{\mu}\left[  \sqrt{-g}f\left(  \phi\right)  F^{\mu\nu}\right]   &
=0,\nonumber
\end{align}
with the energy-momentum tensor
\begin{equation}
T_{\mu\nu}=\partial_{\mu}\phi\partial_{\nu}\phi-\frac{1}{2}g_{\mu\nu}\left(
\partial\phi\right)  ^{2}+f\left(  \phi\right)  \left(  F_{\mu\rho}F_{\nu
}^{\text{ }\rho}-\frac{1}{4}g_{\mu\nu}F_{\rho\sigma}F^{\rho\sigma}\right)  .
\end{equation}
To allow for spontaneous scalarization of KN black holes, the KN black hole
solution with $\phi=0$ must satisfy the equations of motion $\left(
\ref{eq:nonlinearEOMs}\right)  $. This requirement imposes a condition on the
coupling function $f\left(  \phi\right)  $, specifically $f^{\prime}\left(
0\right)  =\left.  df\left(  \phi\right)  /d\phi\right\vert _{\phi=0}=0$.
Without loss of generality, we set $f\left(  0\right)  =1$. Consequently,
$f\left(  \phi\right)  $ can be expanded around $\phi=0$ as%
\begin{equation}
f\left(  \phi\right)  =1+\alpha\phi^{2}+\mathcal{O}\left(  \phi^{3}\right)  ,
\label{eq:se}%
\end{equation}
where $\alpha$ is a dimensionless coupling constant governing the
scalar-electromagnetic interaction strength.

In the EMS models, KN black holes exhibit stability against metric and vector
perturbations, similar to the Einstein-Maxwell theory \cite{Dias:2015wqa}.
However, they may develop a tachyonic instability when subjected to scalar
perturbation $\delta\phi$, leading to the formation of scalarized black holes.
Linearizing the scalar field equation in the KN black hole background yields
\begin{equation}
\left(  \square-\mu_{\text{eff}}^{2}\right)  \delta\phi=0,
\label{eq:linear equation}%
\end{equation}
where the effective mass squared is given by $\mu_{\text{eff}}^{2}=\alpha
F^{\mu\nu}F_{\mu\nu}$. In the Boyer-Linquist coordinates, for a KN black hole
with ADM mass $M$, angular momentum $J$ and electric charge $Q$, the effective
mass squared is given by%
\begin{equation}
\mu_{\text{eff}}^{2}=-\frac{\alpha Q^{2}\left(  r^{4}-6a^{2}r^{2}\cos
^{2}\theta+a^{4}\cos^{4}\theta\right)  }{\left(  r^{2}+a^{2}\cos^{2}%
\theta\right)  ^{4}}, \label{eq:effmass}%
\end{equation}
where $a=J/M$ is the ratio of angular momentum to mass. A tachyonic
instability could arise when $\mu_{\text{eff}}^{2}<0$, potentially driving the
system away from KN black hole solutions. For $\alpha>0$, regions where
$\mu_{\text{eff}}^{2}<0$ consistently appear outside the event horizon in KN
black holes, although these regions shrink as the black hole spin increases
\cite{Guo:2024bkw}. Conversely, when $\alpha<0$, the $\mu_{\text{eff}}^{2}<0$
regions emerge only when the black hole spin is sufficiently large
\cite{Guo:2024lck}.

However, the condition $\mu_{\text{eff}}^{2}<0$ is only a necessary condition
for the appearance of tachyonic instability. To overcome dissipation through
the event horizon and spatial infinity, $\mu_{\text{eff}}^{2}$ must be
sufficiently negative to induce this instability. In other words, only KN
black holes for which $\mu_{\text{eff}}^{2}$ falls below certain threshold
values can develop a tachyonic instability. At these thresholds, the tachyonic
instability triggers the formation of stationary scalar clouds---regular
bound-state solutions to Eq. $\left(  \ref{eq:linear equation}\right)
$---that exist outside KN black holes. These scalar clouds mark bifurcation
points in the parameter space and signal the onset of scalarized KN black
holes. The existence domains for scalar clouds at both fundamental and excited
states have been identified for $\alpha>0$ \cite{Guo:2024bkw} and $\alpha<0$
\cite{Guo:2024lck}, respectively. It is important to note that the existence
domains of these scalar clouds are independent of the specific form of the
coupling function $f\left(  \phi\right)  $, provided it satisfies the series
expansion given in Eq. $\left(  \ref{eq:se}\right)  $.

\subsection{Rotating Black Hole Solutions}

To construct scalarized KN black hole solutions, we employ a generic ansatz
for stationary, axisymmetric and asymptotically-flat black hole solutions
\cite{Herdeiro:2015gia,Delgado:2016jxq,Herdeiro:2020wei,Guo:2023mda}:%
\begin{align}
ds^{2}  &  =-e^{2F_{0}}Ndt^{2}+e^{2F_{1}}\left(  \frac{dr^{2}}{N}+r^{2}%
d\theta^{2}\right)  +e^{2F_{2}}r^{2}\sin^{2}\theta\left(  d\varphi^{2}%
-\frac{W}{r^{2}}dt\right)  ^{2},\nonumber\\
A_{\mu}dx^{\mu}  &  =\left(  A_{t}-A_{\varphi}\frac{W}{r^{2}}\sin
\theta\right)  dt+A_{\varphi}\sin\theta d\varphi\text{ and }\phi=\phi\left(
r,\theta\right)  . \label{eq:ansatz}%
\end{align}
Here $N\equiv1-r_{H}/r$, where $r_{H}$ is the black hole horizon radius. The
seven functions $F_{0}$, $F_{1}$, $F_{2}$, $W$, $A_{t}$, $A_{\varphi}$ and
$\phi$ depend only on the coordinates $r$ and $\theta$.

In the stationary spacetime, two Killing vectors $\partial_{t}$ and
$\partial_{\varphi}$ are present. Their linear combination $\xi=\partial
_{t}+\Omega_{H}\partial\varphi$, where $\Omega_{H}$ is the angular velocity of
the black hole horizon, is null and orthogonal to the horizon. The surface
gravity $\kappa$ is then defined by $\kappa^{2}=-\left(  \nabla_{\mu}\xi_{\nu
}\right)  \left(  \nabla^{\mu}\xi^{\nu}\right)  /2$ and related to the Hawking
temperature $T_{H}$ as \cite{Herdeiro:2015gia}
\begin{equation}
T_{H}=\frac{\kappa}{2\pi}=\frac{1}{4\pi r_{H}}e^{F_{0}\left(  r_{H}%
,\theta\right)  -F_{1}\left(  r_{H},\theta\right)  }. \label{eq:TH}%
\end{equation}
In the EMS models, the black hole entropy is expressed as $S=A_{H}/4$, where
the area of the horizon $A_{H}$ is given by
\begin{equation}
A_{H}=2\pi r_{H}^{2}\int_{0}^{\pi}d\theta\sin\theta e^{F_{1}\left(
r_{H},\theta\right)  +F_{2}\left(  r_{H},\theta\right)  }. \label{eq:Ah}%
\end{equation}

Various physical quantities, including the black hole mass $M$, charge $Q$,
angular momentum $J$, electrostatic potential $\Phi$ and horizon angular
velocity $\Omega_{H}$, can be extracted by analyzing the asymptotic behavior
of the metric and gauge field functions near the horizon and at spatial
infinity \cite{Herdeiro:2015gia,Delgado:2016jxq}:
\begin{align}
\left.  A_{t}\right\vert _{r=r_{H}}  &  \sim0,\quad\left.  W\right\vert
_{r=r_{H}}\sim r_{H}^{2}\Omega_{H},\nonumber\\
\left.  A_{t}\right\vert _{r=\infty}  &  \sim\Phi-\frac{Q}{r},\quad\left.
W\right\vert _{r=\infty}\sim\frac{2J}{r},\quad\left.  e^{2F_{0}}N\right\vert
_{r=\infty}\sim1-\frac{2M}{r}. \label{eq:asymptotic behaviors}%
\end{align}
These quantities further satisfy the Smarr relation
\cite{Herdeiro:2015gia,Guo:2021zed,Fernandes:2022gde}:
\begin{equation}
M=2T_{H}S+2\Omega_{H}J+\Phi Q, \label{eq:smarr}%
\end{equation}
which allows us to assess the accuracy of our numerical solutions.

To obtain scalarized black hole solutions, we numerically solve the coupled
partial differential equations derived by substituting the ansatz in Eq.
$\left(  \ref{eq:ansatz}\right)  $ into the equations of motion $\left(
\ref{eq:nonlinearEOMs}\right)  $. For numerical implementation, we compactify
the radial coordinate $r$ via the transformation%
\begin{equation}
x=\frac{\sqrt{r^{2}-r_{H}^{2}}-r_{H}}{\sqrt{r^{2}-r_{H}^{2}}+r_{H}},
\end{equation}
which maps the event horizon $r=r_{H}$ and spatial infinity $r=\infty$ to
$x=-1$ and $x=1$. Using this compactified coordinate $x$, the power series
expansions near the horizon yield the following boundary conditions at
$x=-1$:
\begin{equation}
\partial_{x}F_{0}=\partial_{x}F_{1}=\partial_{x}F_{2}=\partial_{x}%
\phi=\partial_{x}A_{\varphi}=A_{t}=W-\Omega_{H}=0. \label{eq:bdx0}%
\end{equation}
Meanwhile, boundary conditions at $x=1$ are obtained by imposing flatness at
spatial infinity:
\begin{equation}
F_{0}=F_{1}=F_{2}=\phi=A_{\varphi}=A_{t}-\Phi=W=0. \label{eq:bdx1}%
\end{equation}
On the symmetric axis, axial symmetry and regularity impose the following
conditions at $\theta=0$ and $\theta=\pi$:
\begin{equation}
\partial_{\theta}F_{0}=\partial_{\theta}F_{1}=\partial_{\theta}F_{2}%
=\partial_{\theta}\phi=\partial_{\theta}A_{\varphi}=\partial_{\theta}%
A_{t}=\partial_{\theta}W=0. \label{eq:bdtheta0}%
\end{equation}

In this work, we focus on solutions exhibiting equatorial-plane symmetry,
allowing us to restrict the computational domain to the upper half-plane
$0\leq\theta\leq\pi/2$. Consequently, the $\theta=\pi$ boundary condition in
Eq. $\left(  \ref{eq:bdtheta0}\right)  $ is replaced with
\begin{equation}
\partial_{\theta}F_{0}=\partial_{\theta}F_{1}=\partial_{\theta}F_{2}%
=\partial_{\theta}\phi=\partial_{\theta}A_{\varphi}=\partial_{\theta}%
A_{t}=\partial_{\theta}W=0\text{ at }\theta=\pi/2. \label{eq:bdtheta1}%
\end{equation}
Consequently, Eqs. $\left(  \ref{eq:bdx0}\right)  $, $\left(  \ref{eq:bdx1}%
\right)  $, $\left(  \ref{eq:bdtheta0}\right)  $ and $\left(
\ref{eq:bdtheta1}\right)  $ serve as the boundary conditions for solving the
partial differential equations. Additionally, the absence of conical
singularities imposes $F_{1}=F_{2}$ on the symmetry axis, which provides an
independent consistency check for our numerical results alongside the Smarr
relation \cite{Herdeiro:2015gia,Herdeiro:2020wei}.

This work employs spectral methods to numerically solve the coupled nonlinear
partial differential equations governing the system. Spectral methods are
well-established and particularly effective for solving nonlinear elliptic
partial differential equations. They approximate solutions using a finite
linear combination of basis functions, thereby transforming the differential
equations into a system of algebraic equations. A key advantage of spectral
methods is their exponential convergence rate with increasing resolution,
which far exceeds the linear or polynomial convergence characteristic of
finite difference or finite element methods.

In our numerical work, we implement spectral methods by approximating the
functions of interest, collectively denoted by $\mathcal{F}=\left\{
F_{0},F_{1},F_{2},W,A_{t},A_{\varphi},\phi\right\}  $, as a finite linear
combination of basis functions:
\begin{equation}
\mathcal{F}^{\left(  k\right)  }=\sum_{i=0}^{N_{x}-1}\sum_{j=0}^{N_{\theta}%
-1}a_{ij}^{\left(  k\right)  }T_{i}\left(  x\right)  \cos\left(
2j\theta\right)  , \label{eq:F}%
\end{equation}
where $T_{i}\left(  x\right)  $ represents the $i$-th Chebyshev polynomial,
$a_{ij}^{\left(  k\right)  }$ are the spectral coefficients, and $N_{x}$ and
$N_{\theta}$ denote the resolutions in the radial and angular coordinates,
respectively. To determine $\alpha_{ij}^{\left(  k\right)  }$, we substitute
the spectral expansions $\left(  \ref{eq:F}\right)  $ into the equations of
motion and then discretize the resulting equations at the Gauss-Chebyshev
points. This process transforms the partial differential equations for
$\mathcal{F}^{\left(  k\right)  }$ into a finite system of algebraic equations
for $a_{ij}^{\left(  k\right)  }$. These algebraic equations are then solved
using the Newton-Raphson method, with the root-finding process carried out
using Mathematica's built-in LinearSolve function.

\section{Numerical Results}

\label{sec:Numerical-Results}

\begin{figure}[ptb]
\includegraphics[width=0.5\textwidth]{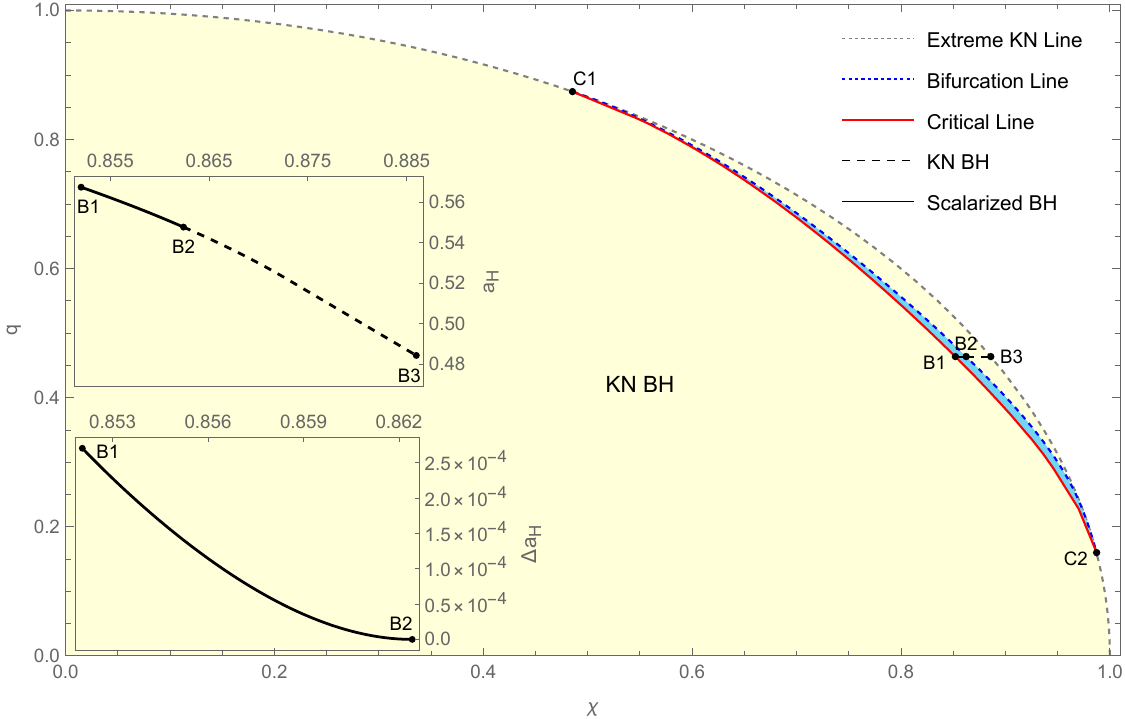}\includegraphics[width=0.5\textwidth]{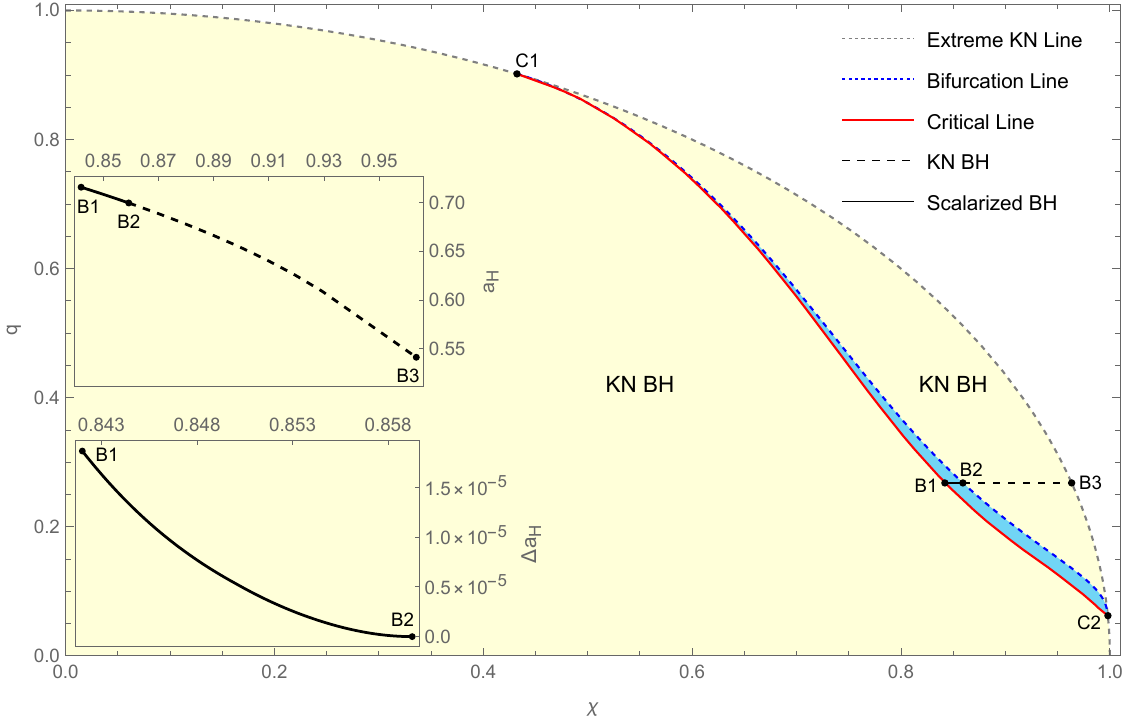}\caption{Domain
of existence for spin-induced scalarized KN black holes in the ($\chi,q$)
parameter space. \textbf{Left Panel: }$\alpha=-100$. \textbf{Right Panel:
}$\alpha=-1000$. The light yellow region indicates the parameter space of KN
black holes, bounded by the extremal KN line corresponding to $q^{2}+\chi
^{2}=1$. The light blue region denotes the domain of existence for scalarized
black holes, within which both KN and scalarized solutions coexist. This
domain is bounded by the bifurcation line (blue dashed), where scalarized
black holes branch off from KN black holes, and the critical line (red solid),
beyond which scalarized black holes can no longer be obtained numerically.
Toward both ends of the existence domain, the extremal KN line, bifurcation
line and critical line converge and merge at the critical points $C_{1}$ and
$C_{2}$. The marked points $B_{1}$, $B_{2}$ and $B_{3}$ on constant-$q$ lines
represent a critical scalarized black hole, a bifurcating KN black hole and an
extremal KN black hole, respectively. The upper inset shows the reduced
horizon area $a_{H}$ as a function of $\chi$ along the constant-$q$ line
connecting $B_{1}$ and $B_{3}$. The lower inset displays $\Delta a_{H}$, the
difference in $a_{H}$ between KN and scalarized black holes, also as a
function of $\chi$. The black dashed line traces KN black hole solutions,
while the black solid line corresponds to scalarized black holes. }%
\label{domain of existence}%
\end{figure}

In this section, we first identify the domain of existence for spin-induced
scalarized KN black holes and subsequently analyze their properties. As
previously discussed, stationary scalar clouds around KN black holes emerge
for $\alpha<0$ when the black hole spin exceeds a critical threshold.
Consequently, we focus on the $\alpha<0$ regime to construct spin-induced
scalarized solutions. For the coupling function, we employ the exponential
form $f\left(  \phi\right)  =e^{\alpha\phi^{2}}$ throughout this section. For
comparison, a quadratic coupling function $f\left(  \phi\right)  =1+\alpha
\phi^{2}$ is examined in Appendix
\ref{sec:Appendix-B:-Domain of Existence for Quadratic Coupling Function}.
Additionally, we restrict our attention to fundamental black hole solutions,
characterized by a nodeless scalar field.

In Appendix \ref{sec:Appendix-A:-Convergence Results for Spectral Method}, we
conduct convergence tests of scalarized black hole solutions by monitoring the
Smarr relation and the absence of conical singularities for varying
resolutions, $N_{x}$ or $N_{\theta}$. As anticipated, the convergence tests
demonstrate exponential convergence as the resolution increases, persisting
until a roundoff plateau is reached. To ensure numerical precision and
efficiency, we employ spectral methods with resolutions $N_{x}=40$ and
$N_{\theta}=11$ to solve the partial differential equations. With these
resolutions, our results indicate that the numerical error of scalarized black
hole solutions is less than $10^{-8}$ when they are sufficiently distant from
the critical line. However, as we approach the critical line, the solutions
exhibit a numerical error on the order of $10^{-5}$. For convenience, we
introduce dimensionless reduced quantities $q\equiv Q/M$, $\chi\equiv J/M^{2}$
and $a_{H}\equiv A_{H}/16\pi M$ in the remainder of this section.

Fig. \ref{domain of existence} displays the domain of existence for
spin-induced scalarized KN black holes in the $\left(  \chi,q\right)  $
parameter space, with $\alpha=-100$ (left panel) and $\alpha=-1000$ (right
panel). The light blue region, bounded by the bifurcation and critical lines,
marks the parameter space where scalarized black holes exist. The bifurcation
line corresponds to the threshold at which the tachyonic instability triggers
the formation of stationary scalar cloud around KN black holes. Both
endpoints, $C_{1}$ and $C_{2}$, of the bifurcation line lie on the extremal KN
black hole line, defined by $q^{2}+\chi^{2}=1$. Starting from the bifurcation
line, scalarized black hole solutions are computed along constant-$\chi$ lines
by varying $q$ until reaching the critical line. Beyond this line, numerical
solutions cannot be reliably obtained with an error below $10^{-5}$.
Interestingly, our numerical results show that the critical line
asymptotically approaches both endpoints of the bifurcation line, implying
that the bifurcation and critical lines merge at $C_{1}$ and $C_{2}$. This
indicates that spin-induced scalarized KN black holes do not exist when the
spin is either too low or too high. Furthermore, Fig.
\ref{domain of existence} illustrates that the existence domain expands as
$\left\vert \alpha\right\vert $ increases.

\begin{figure}[ptb]
\includegraphics[width=0.5\textwidth]{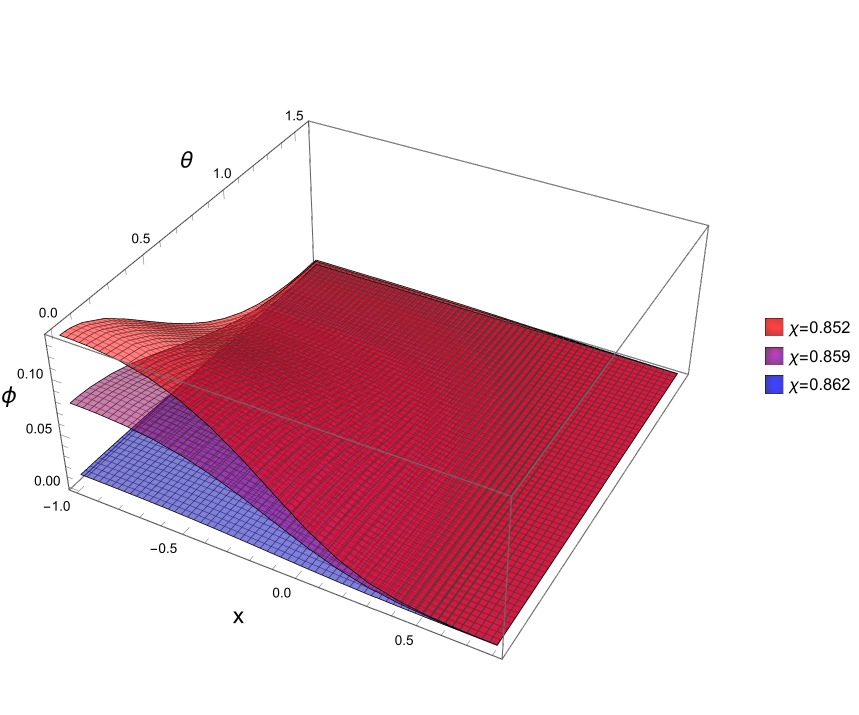}\includegraphics[width=0.5\textwidth]{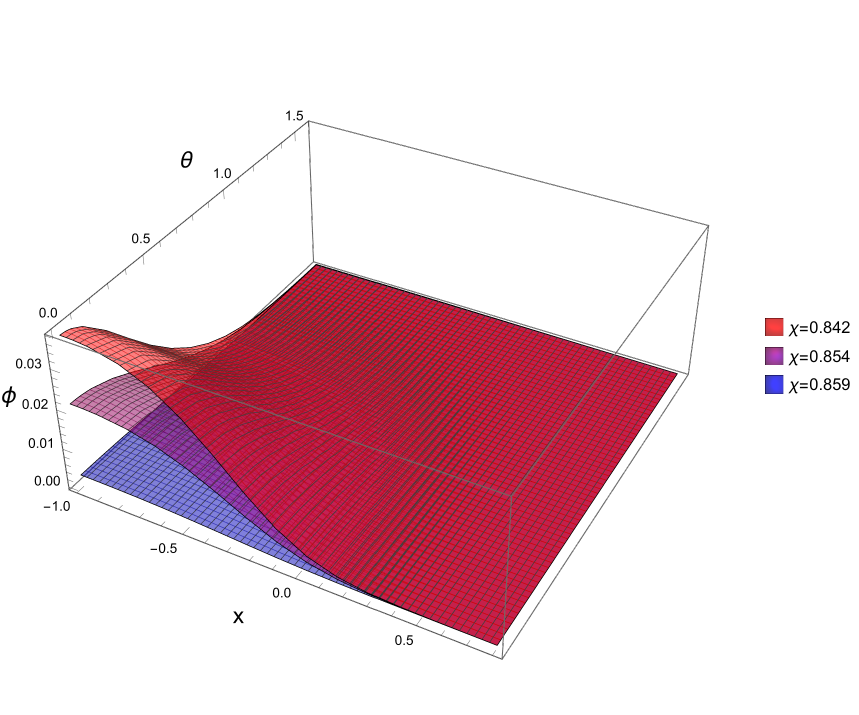}\caption{Scalar
field wave function $\phi\left(  x,\theta\right)  $\textbf{ }for
representative scalarized black holes along the black solid lines in Fig.
\ref{domain of existence} for\textbf{ }$\alpha=-100$ (\textbf{Left Panel}) and
$\alpha=-1000$ (\textbf{Right Panel}). The scalar field corresponding to the
critical scalarized black hole at point $B_{1}$ is highlighted in red and
exhibits the largest amplitude. As the black hole solution approaches the
bifurcation line, the scalar field amplitude decreases. In all cases, the
scalar field wave functions are concentrated near the black hole poles and
decay with increasing distance from the event horizon.}%
\label{scalar field}%
\end{figure}

As shown in \cite{Guo:2024lck}, KN black holes below the bifurcation line are
stable against the scalar perturbation, while those above exhibit a tachyonic
instability. Therefore, Fig. \ref{domain of existence} demonstrates that
spin-induced scalarized KN black holes coexist with stable KN black holes
possessing the same $q$ and $\chi$ within their domain of existence. To
examine the behavior of the horizon area, we highlight three distinct black
hole configurations with the same $q$ in Fig. \ref{domain of existence}:
$B_{1}$, a critical scalarized black hole; $B_{2}$, a black hole on the
bifurcation line; $B_{3}$, an extremal KN black hole. The upper insets show
that the reduced horizon area $a_{H}$ of scalarized black holes decreases as
the spin increases for fixed $q$. The lower insets display $\Delta a_{H}$, the
difference in reduced horizon area between KN and scalarized black holes,
along the constant-$q$ line connecting $B_{1}$ and $B_{2}$. These results
indicate that KN black holes generally have a larger reduced horizon area than
scalarized black holes with the same $q$ and $\chi$, suggesting that KN black
holes are slightly entropically favored over scalarized black holes. In
contrast, for $\alpha>0$, scalarized KN black holes coexist with unstable KN
black holes and are always entropically favored within the coexistence region
\cite{Guo:2023mda}.

Fig. \ref{scalar field} presents representative scalar wave functions
$\phi\left(  x,\theta\right)  $ for $\alpha=-100$ and $-1000$. Three
scalarized black hole solutions along the constant-$q$ line connecting $B_{1}$
and $B_{2}$, are selected. All scalar wave functions exhibit maxima near the
black hole poles and decay outward from the event horizon, consistent with the
fundamental scalar cloud wave functions reported in \cite{Guo:2024lck}.
Notably, rapidly rotating KN black holes with $\alpha>0$ develop
equatorial-plane scalar cloud concentrations \cite{Guo:2024bkw}, contrasting
the polar concentrations observed here. As the black hole spin decreases
(i.e., approaching the critical line), scalar field amplitudes intensify near
the horizon and poles. The scalarized black hole on the critical line exhibits
the largest scalar field amplitude. Additionally, the amplitude of the scalar
field is significantly higher for $\alpha=-100$ compared to $\alpha=-1000$.

\begin{figure}[ptb]
\includegraphics[width=0.5\textwidth]{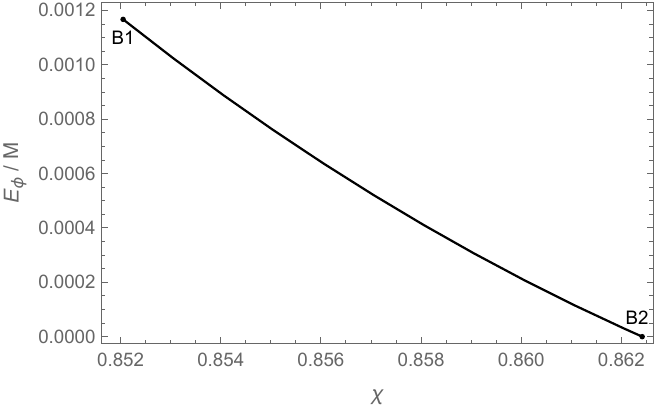}\includegraphics[width=0.5\textwidth]{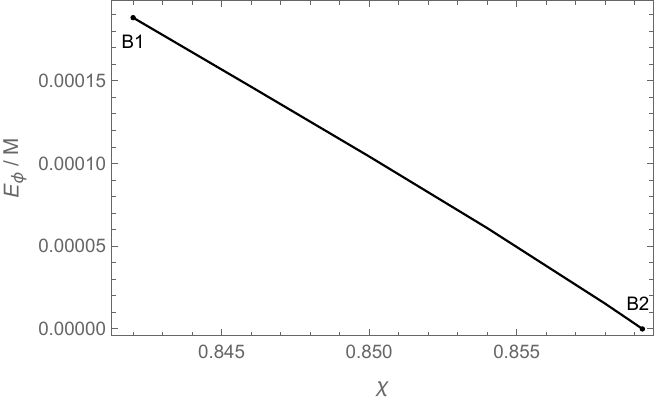}\caption{Ratio
of scalar field energy to black hole mass, $E_{\phi}/M$, as a function of
$\chi$ along the black solid lines in Fig. \ref{domain of existence}.
\textbf{Left Panel: }$\alpha=-100$. \textbf{Right Panel: }$\alpha=-1000$. The
critical scalarized black hole at point $B_{1}$ stores the largest amount of
energy in the scalar field, while the black hole on the bifurcation line at
point $B_{2}$ has zero scalar field energy.}%
\label{scalar field energy}%
\end{figure}

To quantify the energy stored in the scalar field outside the event horizon,
we define
\begin{equation}
E_{\phi}=\int_{r_{H}}^{\infty}n^{\mu}n^{\nu}T_{\mu\nu}^{\phi}dV,
\end{equation}
where $T_{\mu\nu}^{\phi}=\partial_{\mu}\phi\partial_{\nu}\phi-g_{\mu\nu
}\left(  \partial\phi\right)  ^{2}/2$ is the stress-energy tensor of the
scalar field, and $n^{\alpha}$ is the unit normal vector to spatial
hypersurfaces in the $3+1$ decomposition of spacetime. In Fig.
\ref{scalar field energy}, we present $E_{\phi}$ as a function of $\chi$ for
scalarized black holes on the constant-$q$ line connecting $B_{1}$ to $B_{2}$,
for $\alpha=-100$ and $-1000$. The results show that the scalar field energy
increases as $\chi$ decreases (i.e., approaching the critical line).
Furthermore, $E_{\phi}$ constitutes only a small fraction of the black hole
mass $M$, on the order of $10^{-3}$ for $\alpha=-100$ and $10^{-4}$ for
$\alpha=-1000$. This suggests that the backreaction of the scalar field on the
spacetime geometry is limited. Consequently, scalarized KN black hole
solutions do not significantly deviate from KN black holes with scalar cloud
along the bifurcation line. This limited deviation may help explain the
considerably smaller domain of existence compared to cases with positive
coupling. Moreover, the suppressed nonlinear effects of the scalar field also
imply that terms beyond quadratic order in the series expansion of $f\left(
\phi\right)  $, given in Eq. $\left(  \ref{eq:se}\right)  $, do not contribute
significantly. This, in turn, suggests similarities in the features of
scalarized black holes across different coupling functions. Indeed, Appendix
\ref{sec:Appendix-B:-Domain of Existence for Quadratic Coupling Function}
demonstrates that the existence domains for scalarized black holes are quite
similar for both exponential and quadratic coupling functions.

Compared to scalarized KN black holes with $\alpha>0$ discussed in
\cite{Guo:2023mda}, the $\alpha<0$ case exhibits two notable distinctions: (1)
the scalar field energy is substantially smaller, and (2) scalarized KN black
holes coexist with stable KN black holes, rather than unstable ones. To
understand these differences, consider the dynamical evolution of an unstable
KN black hole undergoing a tachyonic instability, and assume that the end
state of this evolution is a scalarized KN black hole$^{\ref{ft:1}}$.
\footnotetext[1]{\label{ft:1} Other final states of the dynamical evolution
are also possible. For instance, an unstable KN black hole may evolve into a
spun-down, stable KN black hole surrounded by a scalar cloud, which is
eventually depleted through electromagnetic and gravitational wave emission or
other dissipative processes.} During the evolution, the scalar field
accumulates outside the event horizon, while scalar, electromagnetic and
gravitational radiation carry away energy and angular momentum to infinity.
Since the scalar field is electrically neutral, the total charge of the system
remains conserved---an observation supported by fully nonlinear simulations of
Reissner-Nordstr\"{o}m (RN) black holes evolving into scalarized RN black
holes \cite{Garcia-Saenz:2025rbc}. Furthermore, it has been shown that slower
spin enhances (suppresses) the tachyonic instability for the $\alpha>0$
$\left(  \alpha<0\right)  $ case \cite{Guo:2024lck,Guo:2024bkw}. At least
during the early stages of the evolution, the spacetime remains well
approximated by a KN black hole. Consequently, in the $\alpha>0$ case, angular
momentum loss accelerates scalar field condensation, whereas in the $\alpha<0$
case, it hinders the process. As a result, the final scalarized KN black hole
in the $\alpha<0$ scenario is expected to deviate less from the initial KN
black hole, which may account for the significantly lower scalar field energy.
Moreover, for $\alpha<0$, if the final scalarized KN black hole coexists with
a KN black hole having the same global charges, the coexisting KN black hole
possesses less angular momentum---and therefore a weaker tachyonic
instability---than the initial (unstable) KN black hole. This reduction in
angular momentum may stabilize the coexisting KN black hole, explaining why
scalarized KN black holes with $\alpha<0$ can coexist with stable KN black holes.

\section{Conclusions}

\label{sec:Conclusions}

In this work, we have investigated the scalarization of KN black holes within
the EMS models, which exhibit a spin-induced tachyonic instability for
negative coupling constants, as identified in \cite{Lai:2022spn,Hod:2022txa}.
By numerically constructing spin-induced scalarized KN black hole solutions
for both exponential and quadratic coupling functions, we analyzed their
domain of existence and physical properties in the $\left(  \chi,q\right)  $
parameter space. Our results show that slowly rotating stationary black holes
in these models are well described by the KN metric, whereas rapidly rotating
ones develop scalar hair. Additionally, we found that the scalar field energy
constitutes only a small fraction of the total black hole mass, indicating
suppressed nonlinear effects of the scalar field during scalarization. This
provides a plausible explanation for the relatively narrow domain of existence
of scalarized black hole solutions and the similarity between solutions
obtained with different coupling functions.

Notably, our analysis reveals that spin-induced scalarized KN black holes
coexist with linearly stable KN black holes. This coexistence may be
attributed to the loss of angular momentum during the scalarization of
unstable KN black holes. If scalarized KN black holes represent the final
state of scalarization, then the coexisting KN black holes must have lower
angular momentum than the initial unstable ones. As the spin-induced tachyonic
instability is suppressed at lower spin, the coexisting KN black holes could
be stabilized, provided the angular momentum loss is sufficiently large.
Electromagnetic radiation may play an important role in achieving this loss,
as electromagnetic wave emission is significantly more efficient than
gravitational wave emission in radiating energy and angular momentum when the
black hole charge is comparable to its mass.

Furthermore, in contrast to the EsGB theories where spin-induced scalarized
black holes are entropically favored, we found that, in the EMS models, KN
black holes possess higher entropy in the coexisting region. Although entropic
preference does not necessarily align with dynamical stability due to
dissipative processes, this observation suggests that spin-induced scalarized
KN black holes might be metastable and could eventually decay into stable KN
black holes.

To better understand these phenomena, future investigations should address two
open questions: (1) the linear stability of spin-induced scalarized KN black
holes against radial and nonaxisymmetric perturbations, and (2) the nonlinear
dynamical evolution of spin-induced scalarization. Resolving these issues will
clarify whether spin-induced scalarized KN black holes represent a transient
phase in black hole evolution or a persistent configuration, with implications
for gravitational wave astronomy and high-energy astrophysics.

\begin{acknowledgments}
We are grateful to Yiqian Chen for useful discussions and valuable comments.
This work is supported in part by NSFC (Grant Nos. 12275183, 12275184,
12347133 and 12250410250).
\end{acknowledgments}

\appendix

\section{Convergence Tests}

\label{sec:Appendix-A:-Convergence Results for Spectral Method}

\begin{figure}[ptb]
\includegraphics[width=1\textwidth]{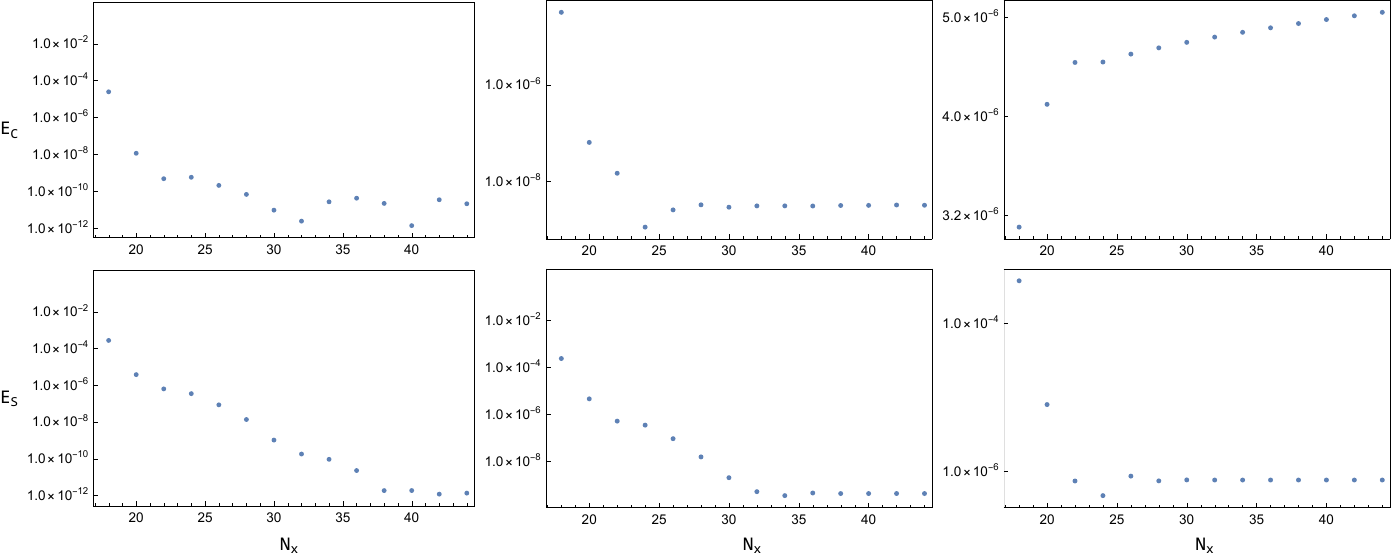}\caption{Logarithmic plots of
the conical singularity error $E_{C}$ (\textbf{Upper Row}) and the Smarr
relation error $E_{S}$ (\textbf{Lower Row}) as functions of the radial
resolution $N_{x}$, with fixed angular resolution $N_{\theta}=11$, for
scalarized black holes near the bifurcation line (\textbf{Left Column}),
between the bifurcation and critical lines (\textbf{Middle Column}) and on the
critical line (\textbf{Right Column}).}%
\label{n*11}%
\end{figure}

\begin{figure}[ptb]
\includegraphics[width=1\textwidth]{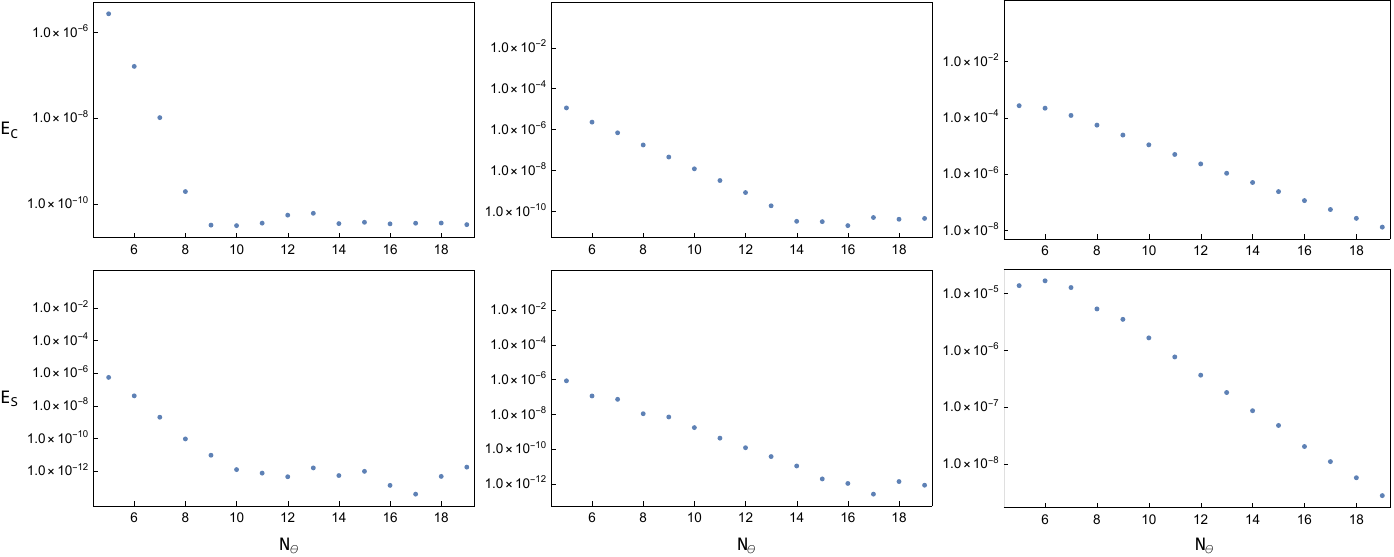}\caption{Logarithmic plots of the
conical singularity error $E_{C}$ (\textbf{Upper Row}) and the Smarr relation
error $E_{S}$ (\textbf{Lower Row}) as functions of the angular resolution
$N_{\theta}$, with fixed radial resolution $N_{x}=40$, for scalarized black
holes near the bifurcation line (\textbf{Left Column}), between the
bifurcation and critical lines (\textbf{Middle Column}) and on the critical
line (\textbf{Right Column}).}%
\label{40*n}%
\end{figure}

To assess the numerical accuracy of our scalarized black hole solutions, we
perform convergence tests by evaluating the absence of conical singularities
and verifying the Smarr relation. Specifically, we define the conical
singularity error as $E_{C}=\underset{-1\leq x\leq1}{\max}|F_{1}\left(
x,0\right)  -F_{2}\left(  x,0\right)  |$ and the Smarr relation error as
$E_{S}=|1-\left(  2T_{H}S+2\Omega_{H}J+\Phi Q\right)  /M|$. Figs. \ref{n*11}
and \ref{40*n} show logarithmic plots of these errors as functions of $N_{x}$
(with $N_{\theta}=11$) and $N_{\theta}$ (with $N_{x}=40$), respectively. The
left, middle and right columns correspond to scalarized black holes near the
bifurcation line, between the bifurcation and critical lines and on the
critical line, respectively. These results demonstrate exponential convergence
of the numerical errors, followed by round-off plateaus$^{\ref{ft:3}}$.
\footnotetext[1]{\label{ft:3} In the upper-right panel of Fig. \ref{n*11}, the
conical singularity error $E_{C}$ for the critical scalarized black hole has
already reached a plateau within the plotted range of the resolution $N_{x}$.}
Notably, the round-off plateau for $E_{S}$ typically occurs later than for
$E_{C}$. In particular, for the scalarized black hole near the bifurcation
line, $E_{S}$ reaches the plateau at $N_{x}=40$ and $N_{\theta}=11$. Based on
these findings, we employ spectral methods with resolutions $N_{x}=40$ and
$N_{\theta}=11$ to solve the partial differential equations, ensuring both
numerical precision and computational efficiency.

\section{Quadratic Coupling Function}

\label{sec:Appendix-B:-Domain of Existence for Quadratic Coupling Function}

\begin{figure}[ptb]
\includegraphics[width=0.5\textwidth]{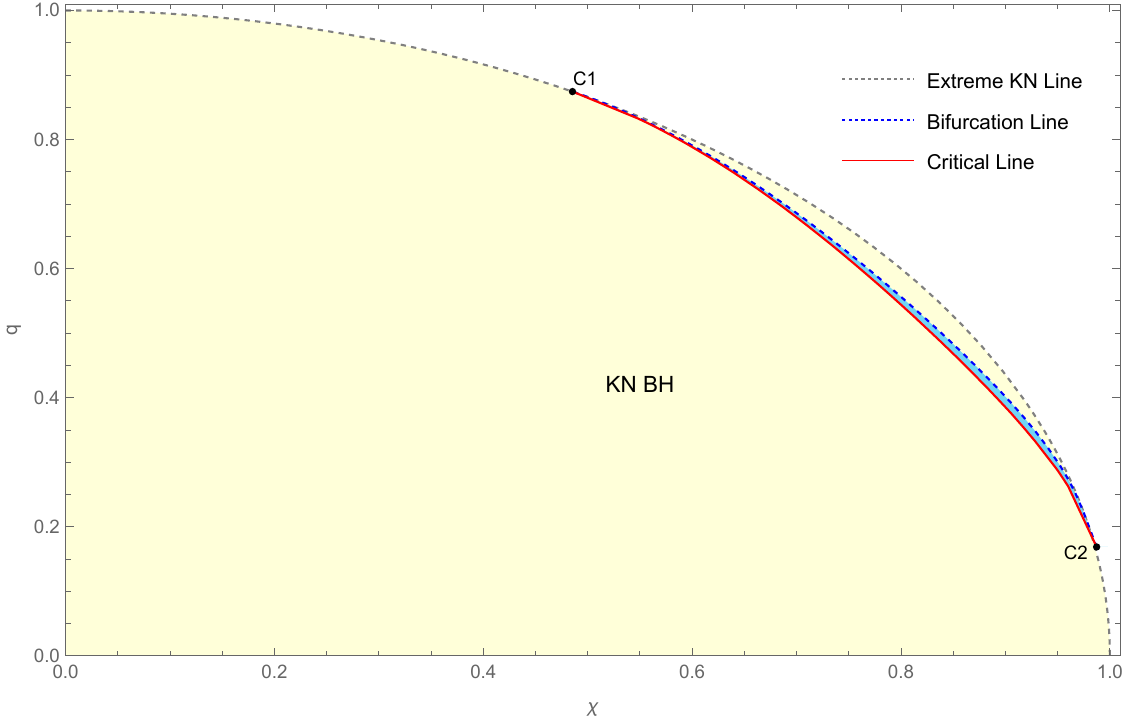}\includegraphics[width=0.5\textwidth]{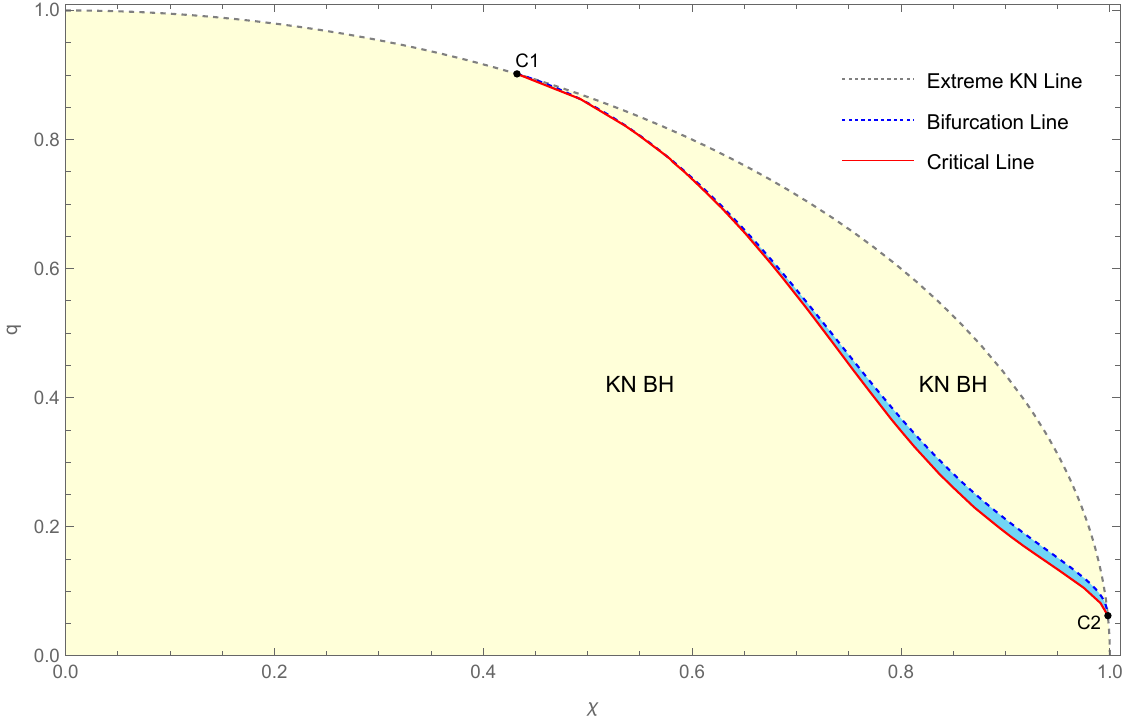}\caption{Domain
of existence for spin-induced scalarized KN black holes in the $\left(
\chi,q\right)  $ plane with the quadratic coupling function $f\left(
\phi\right)  =1+\alpha\phi^{2}$. \textbf{Left Panel: }$\alpha=-100$.
\textbf{Right Panel: }$\alpha=-1000$.}%
\end{figure}

In this appendix, we consider the quadratic coupling function $f\left(
\phi\right)  =1+\alpha\phi^{2}$. The domain of existence for spin-induced
scalarized black hole solutions in this case closely resembles that obtained
with the exponential coupling function, indicating that the specific form of
the coupling function does not significantly affect the structure of
scalarized black hole solutions. As shown in Sec. \ref{sec:Numerical-Results},
the scalar field contributes only a small fraction to the total energy and
therefore plays a limited role in determining the background geometry. The
similarity in the existence domains for different coupling functions may be
attributed to the suppressed nonlinear effects of the scalar field.

\bibliographystyle{unsrturl}
\bibliography{ref}

\end{document}